\documentclass[openacc]{rstransa}

\usepackage{bm}

\begin{document}

\title{Dynamical invariant formalism of shortcuts to adiabaticity}

\author{
Kazutaka Takahashi}

\address{Department of Physics Engineering, Faculty of Engineering,
Mie University, Mie 514--8507, Japan}

\subject{quantum physics, quantum control}



\begin{abstract}
  We give a pedagogical introduction to dynamical invariant formalism of
  shortcuts to adiabaticity.
  For a given operator form of the Hamiltonian with undetermined coefficients,
  the dynamical invariant is introduced to design the coefficients.
  We discuss how the method allows us to realize adiabatic dynamics and
  describe a relation to the counterdiabatic formalism.
  The equation for the dynamical invariant takes a familiar form
  and is often used in various fields of physics.
  We introduce examples of Lax pair, quantum brachistochrone,
  and flow equation.
\end{abstract}


\begin{fmtext}
\end{fmtext}


\maketitle

\section{Introduction}

The method of shortcuts to adiabaticity
is a series of techniques controlling dynamical systems
in efficient ways~\cite{STA13, STA19}.
The word ``adiabaticity'' assumes that
the system is operated very slowly.
However, by using some techniques,
we can mimic adiabatic dynamics under fast operations.

One of the prominent methods is
the counterdiabatic driving~\cite{DR03, DR05, Berry09}.
We introduce an additional term to the time-evolution generator
to prevent nonadiabatic transitions.
The method is reviewed in an article of this issue~\cite{Nakahara22}.

The main aim of this article is to discuss
one of the other useful techniques of shortcuts to adiabaticity.
We discuss a quantity called dynamical invariant~\cite{LR}.
Keeping the original form of the Hamiltonian unchanged, we
can design the protocol so that the state gives a desired time evolution.
The method of control by the dynamical invariant is called 
invariant-based inverse engineering~\cite{CRSCGM}.
The dynamical invariant was also used as a method to treat 
quantum computations~\cite{SDS11}.

The organization of this article is as follows.
First, we discuss fundamental properties of the dynamical invariant
and the idea of inverse engineering in Sec.~\ref{sec:di}.
Next, we give several simple applications in Sec.~\ref{sec:app1}. 
Then, we discuss a relation to the counterdiabatic formalism
in Sec.~\ref{sec:app2} and several applications using
the dynamical invariant in Sec.~\ref{sec:app3}.
The last section \ref{sec:summary} is devoted to summary.

\section{Dynamical invariant formalism}
\label{sec:di}

\subsection{Dynamical invariant}

We consider a quantum system described by
a time-dependent Hamiltonian $H(t)$.
A Hermitian operator $I(t)$ is called
a dynamical invariant or a Lewis--Riesenfeld invariant
when it satisfies 
\begin{align}
 i\hbar\frac{\partial I(t)}{\partial t}=[H(t),I(t)]. \label{di}
\end{align}
As we discuss in the following sections, this type of operators
is familiar in quantum mechanics and is found in many different contexts.
Here, we describe the fundamental properties of the dynamical invariant.

The formal solution of Eq.~(\ref{di}) is written as 
\begin{align}
 I(t)=U(t)I(0)U^\dag(t), \label{u}
\end{align}
where $U(t)$ is the unitary time-evolution operator satisfying 
\begin{align}
 i\hbar\partial_tU(t)=H(t)U(t),
\end{align}
with $U(0)=1$.
Equation (\ref{u}) shows that the eigenvalues of $I(t)$ are
time independent.
We write the spectral representation 
\begin{align}
 I(t)=\sum_n\lambda_n |\phi_n(t)\rangle\langle\phi_n(t)|.
\end{align}
$\{\lambda_n\}_{n=1,2,\dots}$ represents the set of eigenvalues and
each element takes a real constant value.
$\{|\phi_n(t)\rangle\}_{n=1,2,\dots}$ is the corresponding set of eigenstates.
Comparing this spectral representation with Eq.~(\ref{u}), we find that 
$|\phi_n(t)\rangle$ is equivalent to
$|\chi_n(t)\rangle=U(t)|\phi_n(0)\rangle$ up to a phase.
We can write 
\begin{align}
 |\chi_n(t)\rangle=e^{i\alpha_n(t)}|\phi_n(t)\rangle,
\end{align}
where $\alpha_n(t)$ is real.
We apply $i\hbar\partial_t-H(t)$ on both sides of this equation.
Since $|\chi_n(t)\rangle$ satisfies the Schr\"odinger equation, 
we have 
\begin{align}
  0=\left(i\hbar\partial_t-H(t)
  -\hbar\dot{\alpha}(t)\right)|\phi_n(t)\rangle,
\end{align}
where the dot symbol denotes the time derivative.
Then, we obtain
\begin{align}
 \alpha_n(t) =\frac{1}{\hbar}\int_0^t
 \langle\phi_n(s)|(i\hbar\partial_s-H(s))|\phi_n(s)\rangle ds.
 \label{alpha}
\end{align}
We note that $|\chi_n(t)\rangle$ is independent of the phase choice 
of $|\phi_n(t)\rangle$, except the one at the initial time.
$|\chi_n(t)\rangle$ is invariant under the replacement
$|\phi_n(t)\rangle\to e^{i\theta_n(t)}|\phi_n(t)\rangle$
where $\theta_n(t)$ represents
an arbitrary real function with $\theta_n(0)=0$.
The phase $\alpha_n(t)$ takes a familiar form
known in the adiabatic approximation of dynamical systems.
The first term in Eq.~(\ref{alpha}) is known as the geometric phase 
and the second term as the dynamical phase.
In fact, the general solution of the Schr\"odinger equation is written as
\begin{align}
  |\psi(t)\rangle=\sum_n c_n e^{i\alpha_n(t)}|\phi_n(t)\rangle,
\end{align}
where $c_n$ represents a constant determined from the initial condition.
This representation denotes that the solution of the Schr\"odinger equation 
is given by the ``adiabatic state'' of the dynamical invariant.

The term ``invariant'' indicates that the eigenvalues of
$I(t)$ are time independent.
In the classical limit, the commutation relation is replaced by 
the Poisson bracket and the operation in Eq.~(\ref{di}) is interpreted
as the total time derivative: 
\begin{align}
  \frac{d}{dt}(\cdot) =
  \frac{\partial}{\partial t}(\cdot)-\frac{1}{i\hbar}[H,(\cdot)].
\end{align}
Then, the classical analogue of $I(t)$ becomes a constant of motion.

A remarkable feature of this method is that
the time-evolved state can be obtained
by solving the eigenvalue problem of the dynamical invariant.
Of course, 
we must find the explicit operator form of the dynamical invariant
before attacking the eigenvalue problem.
For our aim to realize an ideal control of the system, 
we see in the following that it is not necessarily
to solve the eigenvalue problem.

\subsection{Invariant-based inverse engineering}

The authors in Ref.~\cite{CRSCGM} proposed to use
the dynamical invariant for a control of dynamical systems.
The term ``shortcut to adiabaticity'' was coined there.
In this subsection,
we review the method of inverse engineering briefly.

To understand the fundamental idea, it is convenient to
use a basis-operator representation as discussed in
Refs.\cite{SDS11, GN12, Takahashi13, TMM14}.
We assume that the Hilbert space of the system has finite 
dimension $N$.
We write the Hamiltonian 
\begin{align}
 H(t)=\sum_\mu h_\mu(t)X_\mu,
\end{align}
and the dynamical invariant 
\begin{align}
 I(t)=\sum_\mu b_\mu(t)X_\mu.
\end{align}
Each component of $\{h_\mu(t)\}$ and $\{b_\mu(t)\}$ is real.
Greek indices generally run from 1 to $N^2$.
$\{X_\mu\}$ represents the set of basis operators.
Their operators are Hermitian and satisfy the orthonormal condition 
\begin{align}
 \frac{1}{N}{\rm Tr}\, X_\mu X_\nu=\delta_{\mu,\nu},
\end{align}
and the commutation relation 
\begin{align}
 [X_\mu,X_\nu]=i\sum_\lambda f_{\mu\nu\lambda}X_\lambda,
\end{align}
where $f_{\mu\nu\lambda}$ represents the structure constant.
$f_{\mu\nu\lambda}$ is real and totally antisymmetric
under permutation of indices.
The simplest example is when $N=2$.
Then, the basis operators are given by the unit operator and
the three Pauli operators, for example.

By using the basis-operator representation, we find that
Eq.~(\ref{di}) is written as 
\begin{align}
  \dot{b}_\mu(t)=\sum_{\nu,\lambda}f_{\mu\nu\lambda}h_\nu(t) b_\lambda(t).
  \label{dix}
\end{align}
To obtain the dynamical invariant, we solve this set of equations 
for a given set of $\{h_\mu(t)\}$.
Although these equations are linear ones, it is generally a difficult
task even when the dimension of the Hilbert space is not so large.

In the inverse engineering, as the name suggests,
we obtain $\{h_\mu(t)\}$ for a given set of $\{b_\mu(t)\}$.
Then, Eq.~(\ref{dix}) is interpreted as a simple algebraic relation.
There is no need to solve differential equations.
We can introduce a vector representation 
\begin{align}
  A[b(t)]\left(\begin{array}{c} h_1(t) \\ h_2(t) \\ \vdots
  \end{array}\right)
  =\left(\begin{array}{c} \dot{b}_1(t) \\ \dot{b}_2(t) \\
    \vdots \end{array}\right), \label{div}
\end{align}
where $A$ represents an antisymmetric real matrix.
Each component of $A$ is a linear combination of $b_\mu(t)$.
Since $A$ is not invertible, we must be careful
in handling this matrix equation~\cite{TMM14}.
Here, we do not give general discussions
on the formal solution of Eq.~(\ref{div}).

In principle, the solution $\{h_\mu(t)\}$ in Eq.~(\ref{dix})
can be obtained for various choices of $\{b_\mu(t)\}$.
The original aim of shortcuts to adiabaticity is to prevent nonadiabatic
transitions in systems under control.
Then, the problem is to find appropriate choices of $\{b_\mu(t)\}$
that meet the purposes of system control.

The solution of the original Schr\"odinger equation is
represented by the adiabatic state of the dynamical invariant.
Since the dynamical invariant is not an observable quantity,
this property is not a convenient one.
When we consider a time evolution from $t=0$ to $t=t_{\rm f}$,
we set the condition that the dynamical invariant and
the Hamiltonian commute with each other at $t=0$ and $t=t_{\rm f}$: 
\begin{align}
 [H(0),I(0)]=[H(t_{\rm f}),I(t_{\rm f})]=0. \label{bc}
\end{align}
Then, it becomes possible to find a time evolution
from an eigenstate of $H(0)$ to that of $H(t_{\rm f})$.
These conditions can be written as 
\begin{align}
  \sum_{\nu,\lambda}f_{\mu\nu\lambda}h_\nu(0) b_\lambda(0)
  =\sum_{\nu,\lambda}f_{\mu\nu\lambda}h_\nu(t_{\rm f}) b_\lambda(t_{\rm f})=0.
  \label{bcx1}
\end{align}
By using Eq.~(\ref{dix}), we can also write Eq.~(\ref{bcx1}) as 
\begin{align}
 \dot{b}_\mu(0)=\dot{b}_\mu(t_{\rm f})=0,
\end{align}
which means that we start and finish
the time evolution of $b_\mu(t)$ slowly.
In the inverse engineering, we determine $\{h_\mu(t)\}$
for a given $\{b_\mu(t)\}$.
When we choose $\{b_\mu(t)\}$, we must be careful 
for the boundary conditions of $\{b_\mu(t)\}$ 
at $t=0$ and $t=t_{\rm f}$
so that the Hamiltonian at those times can take proper forms.
We treat several examples in the next section.

We summarize the invariant-based inverse engineering as follows.
First, we find an operator form of $I(t)$ satisfying Eq.~(\ref{di})
for a given operator form of $H(t)$.
The time-dependent coefficients of $I(t)$ and $H(t)$
are related with each other.
Second, we choose a specific form of the coefficients of $I(t)$.
Third, the coefficients of $H(t)$
are determined from Eq.~(\ref{dix}).
The coefficients are chosen so that  
the boundary conditions in Eq.~(\ref{bc}) are satisfied.

In practical applications, 
the most important point in the first step is
that $H(t)$ and $I(t)$ are expanded
in terms of a small number of operators.
The equation can always be solved if we use all kinds of operators
$\{X_\mu\}_{\mu=1,2,\dots, N^2}$ defined in the $N$-dimensional Hilbert space.
However, it is not useful when we consider
the implementation of the protocol. 
The exact compact solution
is known for limited cases as we discuss in the following sections.
In the second step, we have many possible choices of
the coefficients of $I(t)$.
They are determined so that $H(t)$ obtained in the third step has
a physically feasible form.
Furthermore, they are required to satisfy
the boundary conditions at $t=0$ and $t=t_{\rm f}$.
These constraints restrict possible forms of
the coefficients significantly.

The advantage of this method is that the original form of
the Hamiltonian is unchanged.
We do not need to introduce additional operators to the Hamiltonian
in contrast with the counterdiabatic driving.
Furthermore, once if we can find a set of operators
satisfying Eq.~(\ref{di}), the procedure becomes simple.
We do not need to solve difficult problems 
such as eigenvalue problems and differential equations.
On the other hand,
finding a possible operator form of the dynamical invariant
becomes a formidable task except several known examples.
We also find a difficulty
when the Hamiltonian is restricted to a specific form.
In that case, even we can find a formal solution of $I(t)$
from Eq.~(\ref{di}),
$\{h_\mu(t)\}$ from Eq.~(\ref{dix}) for a given $\{b_\mu(t)\}$ 
often gives an infeasible form.

\section{Examples}
\label{sec:app1}

\subsection{Two-level system}

As the simplest application, we treat the case where the dimension
of the Hilbert space is equal to two~\cite{CTM11}.
This example is used to drive a single spin-$1/2$ particle.
The magnetic field is applied to control the spin state.

In this case, the standard basis operators are given by the
Pauli operators $\bm{\sigma}=(\sigma^x,\sigma^y,\sigma^z)$.
The Hamiltonian of the system is generally written as 
\begin{align}
  H(t)=\frac{\hbar}{2}h(t)\bm{n}(t)\cdot\bm{\sigma}, 
\end{align}
where $h(t)$ is nonnegative and $\bm{n}(t)$ is a unit vector.
The dynamical invariant is also written
by using a unit vector $\bm{e}(t)$ as 
\begin{align}
  I(t)=\bm{e}(t)\cdot\bm{\sigma}.
\end{align}
The eigenvalues of $I(t)$ are $\pm 1$ since $I^2(t)=1$.
We note that the dynamical invariant generally has
the ambiguity of a multiplicative constant.
The part proportional to the unit operator is an irrelevant constant 
is dropped out without losing generality.
Calculating the commutation relations,
we obtain from Eq.~(\ref{dix}) 
\begin{align}
  \dot{\bm{e}}(t)=h(t)\bm{n}(t)\times\bm{e}(t). \label{div2}
\end{align}

As a simple example, we parametrize $\bm{e}(t)$ as 
\begin{align}
  \bm{e}(t)=\left(\begin{array}{c}
    \sin\theta(t) \\ 0 \\ \cos\theta(t) \end{array}\right).
\end{align}
The time dependence of $\theta(t)$ is determined below.
Solving Eq.~(\ref{div2}) with respect to $\bm{n}(t)$, we obtain 
\begin{align}
 \bm{n}(t)=\left(\begin{array}{c}
    \sqrt{1-\left(\frac{\dot{\theta}(t)}{h(t)}\right)^2}\sin\theta(t) \\
    \frac{\dot{\theta}(t)}{h(t)} \\
    \sqrt{1-\left(\frac{\dot{\theta}(t)}{h(t)}\right)^2}\cos\theta(t)
  \end{array}\right). \label{n}
\end{align}
We see that the difference between $\bm{e}$ and $\bm{n}$ represents
nonadiabatic effects.
When the magnitude of $\dot{\theta}(t)/h(t)$
takes a small value, $\bm{n}$ is close to $\bm{e}$, 
which is consistent with the property that
the adiabaticity condition is determined by
$|\dot{\theta}(t)|/h(t)\ll 1$.

We examine the boundary conditions 
\begin{align}
  \dot{\theta}(0)=\dot{\theta}(t_{\rm f})=0.
\end{align}
There are many possible choices of $\theta(t)$
satisfying these boundary conditions.
The simplest choice is a polynomial function 
\begin{align}
  \theta(t)=\theta(0)+(\theta(t_{\rm f})-\theta(0))
  \left[3\left(\frac{t}{t_{\rm f}}\right)^2
  -2\left(\frac{t}{t_{\rm f}}\right)^3\right]. \label{thetapoly}
\end{align}
It is required that
the resulting $\bm{n}(t)$ takes a physically feasible form.
In the present example, we see that 
the condition $|\dot{\theta}(t)|/h(t)\le 1$ is required.
It gives the relation 
\begin{align}
  h(t)t_{\rm f}\ge 6|\theta(t_{\rm f})-\theta(0)|
  \left[\frac{t}{t_{\rm f}}
    -\left(\frac{t}{t_{\rm f}}\right)^2\right]. \label{threshold}
\end{align}
We need to take a large value of $h(t)t_{\rm f}$ so that 
this relation holds for any $t$ with $0\le t\le t_{\rm f}$.

\begin{figure}[t]
\centering\includegraphics[width=0.8\columnwidth]{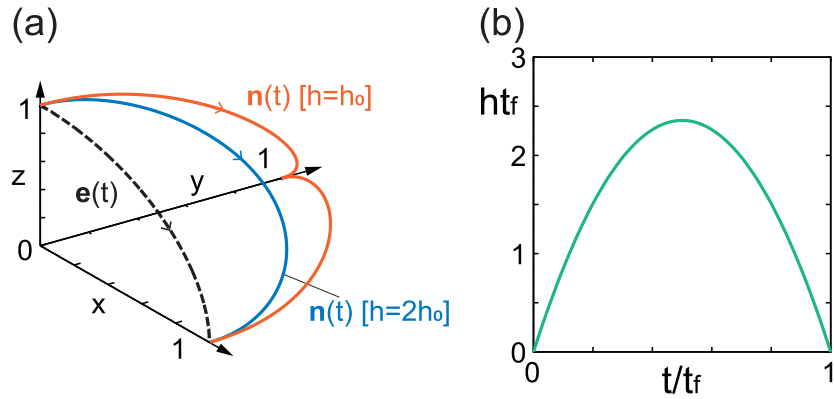}
\caption{
  Inverse engineering for a two-level system.
  (a). A trajectory of $\bm{e}(t)$ for the dynamical invariant  
  is denoted by the dashed (black) line.
  The corresponding trajectories of $\bm{n}(t)$
  are denoted by solid lines.
  We take $h=2h_0$ (blue) and $h=h_0$ (red) where 
  $h_0=3\pi/4t_{\rm f}$ represents the threshold value
  determined from Eq.~(\ref{threshold}).
  We note that all the trajectories are on the unit sphere.
  (b). The direction of the magnetic field is fixed to
  the $y$-direction as $\bm{n}(t)=(0,1,0)$.
  We plot the magnetic field $h(t)$ to be applied
  when $\bm{e}(t)$ takes a trajectory presented in the panel (a).
}
\label{fig1}
\end{figure}

We show a trajectory of $\bm{e}(t)$ and 
the corresponding $\bm{n}(t)$ in the panel (a) of Fig.~\ref{fig1}.
We consider the case
\begin{align}
  \theta(0)=0, \quad \theta(t_{\rm f})=\frac{\pi}{2},
\end{align}
and set $h(t)$ to a time-independent value.
We note that $\bm{e}(t)$ represents the Bloch vector
and $\bm{n}(t)$ is the direction of the magnetic field to be applied.
We find a singular behavior of $\bm{n}(t)$ at the point
where the equality holds in Eq.~(\ref{threshold}).
The adiabaticity condition $h(t)t_{\rm f}\gg 1$
is required to obtain a smooth trajectory
close to $\bm{e}(t)$.

It is also possible to keep the vector $\bm{n}(t)$
in the $y$-direction.
We choose the magnitude of the magnetic field as 
\begin{align}
  h(t)=|\dot{\theta}(t)|=\frac{6|\theta(t_{\rm f})-\theta(0)|}{t_{\rm f}}
  \left[\frac{t}{t_{\rm f}}
  -\left(\frac{t}{t_{\rm f}}\right)^2\right]. 
\end{align}
This is plotted in the panel (b) of Fig.~\ref{fig1}.
Then, we find from Eq.~(\ref{n}) that $\bm{n}(t)$
is independent of $\theta(t)$.
The magnetic field is applied to the axis 
perpendicular to the plane where the Bloch vector lies.
This protocol is easily understood without using the present technique.

When the Hamiltonian takes a restricted form, 
finding $b_\mu(t)$, $\bm{e}(t)$ in the present example,
with the required boundary conditions
becomes a cumbersome task. 
For example, one of the components of $\bm{n}(t)$ is set to zero: 
\begin{align}
  \bm{n}(t)=\left(\begin{array}{c}
    \sin\Theta(t) \\ 0 \\ \cos\Theta(t) \end{array}\right).
\end{align}
Then, parametrizing $\bm{e}(t)$ as 
\begin{align}
  \bm{e}(t)=\left(\begin{array}{c}
    \sin\theta(t)\cos\varphi(t) \\ \sin\theta(t)\sin\varphi(t) \\
    \cos\theta(t) \end{array}\right),
\end{align}
we obtain the relation between $h(t)\bm{n}(t)$ and $\bm{e}(t)$ 
\begin{align}
  h(t)\left(\begin{array}{c} \cos\Theta(t) \\ \sin\Theta(t)
  \end{array}\right)
  =\left(\begin{array}{c}
    -\frac{\dot{\theta}(t)}{\tan\theta(t)\tan\varphi(t)}+\dot{\varphi}(t) \\
    -\frac{\dot{\theta}(t)}{\sin\varphi(t)}
  \end{array}\right).
\end{align}
Since each component of the right hand side goes to infinity
at $\theta\to 0$ and $\varphi\to 0$, a careful choice is
required for $\bm{e}(t)$. 

In the above examples, we only discussed pure state systems.
It is a straightforward task to apply the formalism to mixed states.
We can find some application in Ref.~\cite{FWN12}.

A similar analysis is possible when the dimension of the Hilbert space
is not so large.
Four-level systems were discussed in Refs.~\cite{GN12, GWN14}
based on the Lie algebraic structure.
We can find applications of few-level systems
under various settings in many works~\cite{STA13, STA19}.
The result for two-level systems was also used to describe
many-spin systems with
mean-field interactions~\cite{Takahashi17, Takahashi19}.
Furthermore, a similar analysis is possible for
a generating function of full counting statistics
in a classical stochastic system~\cite{THFH20}.

\subsection{Harmonic oscillator}

In the general discussion and the example of two-level systems,
we treated the case where the dimension of the Hilbert space is finite.
It is possible to apply the same idea to systems with
infinite-dimensional Hilbert space.
We next consider a harmonic oscillator whose angular frequency
changes as a function of time.
This system was first discussed in Ref.~\cite{LR}
to solve the Schr\"odinger equation with
the time-dependent Hamiltonian.
The result was used to implement the inverse engineering
in Ref.~\cite{CRSCGM}.

We consider the one-dimensional Hamiltonian 
\begin{align}
  H(t)=\frac{1}{2m}p^2+\frac{m}{2}\omega^2(t)x^2,
\end{align}
with the position and momentum operators, $x$ and $p$.
The particle mass $m$ represents a positive constant and 
the angular frequency  $\omega(t)$ is a time-dependent function.
Then, it was found in Ref.~\cite{LR} that the following form
of $I(t)$ satisfies Eq.~(\ref{di}):
\begin{align}
  I(t)=\frac{1}{2m}\left(b(t)p-m\dot{b}(t)x\right)^2
  +\frac{m\omega^2(0)}{2}\left(\frac{x}{b(t)}\right)^2, \label{diho}
\end{align}
provided that $b(t)$ obeys the Ermakov equation 
\begin{align}
 \ddot{b}(t)+\omega^2(t)b(t)=\frac{\omega^2(0)}{b^3(t)}. \label{ermakov}
\end{align}
For a given $b(t)$, $\omega(t)$ is determined from the relation 
\begin{align}
  \omega^2(t)=\frac{1}{b(t)}\left(-\ddot{b}(t)
  +\frac{\omega^2(0)}{b^3(t)}\right). \label{omega}
\end{align}
The boundary conditions are given by $\dot{I}(0)=\dot{I}(t_{\rm f})=0$,
which give 
\begin{align}
 \dot{b}(0)=\dot{b}(t_{\rm f})=0, \quad \ddot{b}(0)=\ddot{b}(t_{\rm f})=0.
\end{align}
The simplest polynomial function is 
\begin{align}
  b(t)=1+
  \left(\sqrt{\frac{\omega(0)}{\omega(t_{\rm f})}}-1\right)
  \left[10\left(\frac{t}{t_{\rm f}}\right)^3
  -15\left(\frac{t}{t_{\rm f}}\right)^4
  +6\left(\frac{t}{t_{\rm f}}\right)^5\right]. \label{bho}
\end{align}

\begin{figure}[t]
\centering\includegraphics[width=0.8\columnwidth]{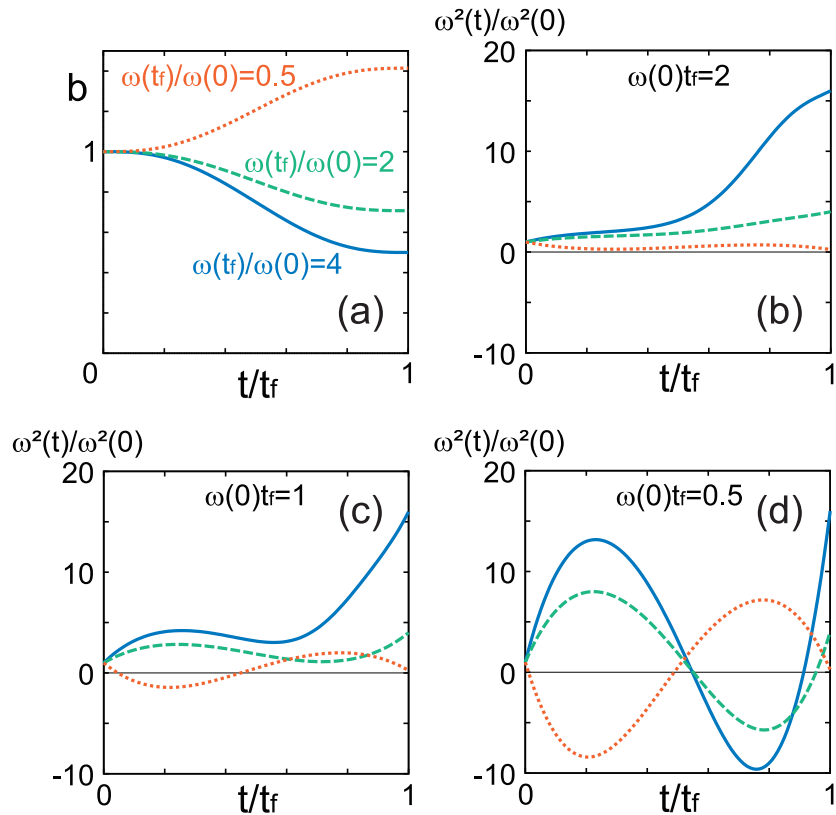}
\caption{
  Inverse engineering for a harmonic oscillator system.
  In the panel (a), we plot $b(t)$ in Eq.~(\ref{bho})
  for several values of $\omega(t_{\rm f})/\omega(0)$.
  The corresponding results of $\omega^2(t)$ from Eq.~(\ref{omega})
  are plotted in the panels (b)-(d).
  We set $\omega(0)t_{\rm f}=2$ in the panel (b),
  $\omega(0)t_{\rm f}=1$ in (c), and 
  $\omega(0)t_{\rm f}=0.5$ in (d).
}
\label{fig2}
\end{figure}

In the panel (a) of Fig.~\ref{fig2},
we plot trajectories of $b(t)$ for several values of
$\omega(t_{\rm f})/\omega(0)$.
The resulting $\omega(t)$ depends not only on $b(t)$
but also on $\omega(0)t_{\rm f}$.
We plot $\omega^2(t)$ in the panels (b)-(d) of Fig.~\ref{fig2}. 
We see that $\omega^2(t)$ strongly depends on the value
of $\omega(0)t_{\rm f}$.
For a large $\omega(0)t_{\rm f}$, the adiabaticity condition is satisfied
and the corresponding result of $\omega(t)$ has a smooth trajectory.
In the opposite limit of small $\omega(0)t_{\rm f}$,
we find that $\omega^2(t)$ shows a rapid change and
goes negative in some cases.
These properties are basically the same as the previous example
of two-level systems.
Generally speaking, the inverse engineering
becomes problematic when the adiabaticity condition is not satisfied.

The described procedure above is enough
to find protocols to be implemented.
As a supplementary calculation,
we demonstrate the diagonalization of 
the dynamical invariant in the present example.
Finding the explicit forms of the eigenstates is instructive
since we can discuss
how the quantum state changes as a function of $t$.
We introduce an operator 
\begin{align}
  a(t)=\sqrt{\frac{m\omega_0}{2\hbar}}\frac{x}{b(t)}
  +\frac{i}{\sqrt{2\hbar m\omega_0}}(b(t)p-m\dot{b}(t)x).
\end{align}
It satisfies the commutation relation $[a(t),a^\dag(t)]=1$ and 
the dynamical invariant is written as 
\begin{align}
  I(t)=\left(a^\dag(t)a(t)+\frac{1}{2}\right)\hbar\omega_0.
\end{align}
Thus, the dynamical invariant can easily be diagonalized by 
the standard procedure of harmonic oscillators.

When we start the time evolution from the ground state of $H(0)$,
the wave function, the solution of the Schr\"odinger equation,
is obtained from $a(t)|\psi(t)\rangle=0$ up to a phase.
It is written in a coordinate representation as 
\begin{align}
  \langle x|\psi(t)\rangle
  =e^{i\alpha(t)}
  \left(\frac{m\omega(0)}{\pi\hbar b^2(t)}\right)^{1/4}
  \exp\left[ -\frac{m\omega(0)}{2\hbar}
    \left(1-i\frac{b(t)\dot{b}(t)}{\omega(0)}\right)
    \left(\frac{x}{b(t)}\right)^2\right].
\end{align}
In the adiabatic approximation, the wave function is represented by
a real Gaussian form except the phase $e^{i\alpha(t)}$.
We see that the nonadiabatic effect in this case
is represented by the imaginary part in the second exponential function.
It gives an oscillating behavior of the wave function.
At $t=0$ and $t=t_{\rm f}$, the imaginary part vanishes and
the wave function coincides with that by the adiabatic approximation.

We note that Eq.~(\ref{diho}) is not the only possible form of
the dynamical invariant.
For example, the following linear form satisfies Eq.~(\ref{di}): 
\begin{align}
  I(t)=b(t)p-m\dot{b}(t)x,
\end{align}
provided that $b(t)$ satisfies the equation
for classical harmonic oscillators 
\begin{align}
  \ddot{b}(t)=-\omega^2(t)b(t).
\end{align}
This linear invariant was discussed in Ref.~\cite{LL82} and 
we can find some application to quantum field theory~\cite{PFR07}.
This solution restricts possible protocols in the inverse engineering
because the boundary conditions $\dot{b}(0)=\dot{b}(t_{\rm f})=0$
and $\ddot{b}(0)=\ddot{b}(t_{\rm f})=0$
give $\omega(0)=\omega(t_{\rm f})=0$.
This linear invariant was used 
for momentum or position scaling~\cite{MMPPT20} 
and for coupled/multi-dimensional
harmonic oscillators~\cite{TTLPM20,SM21,LLM22}.

The harmonic oscillator Hamiltonian only involves quadratic operators 
and we can construct a dynamical invariant 
which is a homogeneous polynomial of $x$ and $p$.
This property is due to commutation relations 
\begin{align}
  &[(\mbox{$k$th-order polynomial of $x$ and $p$}),
    (\mbox{quadratic polynomial of $x$ and $p$})] \nonumber\\
  &=
  (\mbox{$k$th-order polynomial of $x$ and $p$}).
\end{align}
We can construct a closed algebra within a limited space of operators.
We note a general property of the dynamical invariant 
that the product of dynamical invariants
also represents a dynamical invariant.
It is not evident whether we can find higher-order invariants
that cannot be factorized.

As a nontrivial generalization,
it is known that the following set of operators
satisfy Eq.~(\ref{di}):
\begin{align}
  & H(t)=\frac{1}{2m}p^2-F(t)x+\frac{m}{2}\omega^2(t)x^2
  +\frac{1}{b^2(t)}U\left(\frac{x-x_{\rm c}(t)}{b(t)}\right), \\
  & I(t) =\frac{1}{2m}\left[b(t)(p-m\dot{x}_{\rm c}(t))
    -m\dot{b}(t)(x-x_{\rm c}(t))\right]^2 
  +\frac{m\omega^2(0)}{2}\left(\frac{x-x_{\rm c}(t)}{b(t)}\right)^2
  +U\left(\frac{x-x_{\rm c}(t)}{b(t)}\right), 
\end{align}
where $F(t)$ and $U(x)$ represent arbitrary functions.
Here, $b(t)$ satisfies the Ermakov equation (\ref{ermakov}) and
$x_{\rm c}(t)$ satisfies the equation for a forced oscillator:
\begin{align}
  \ddot{x}_{\rm c}(t)+\omega^2(t)x_{\rm c}(t)=\frac{1}{m}F(t).
\end{align}
This type of the Hamiltonian was discussed in Ref.~\cite{LL82} and
was used for an inverse engineering in Ref.~\cite{TICRGM11}.
The potential $U((x-x_{\rm c}(t))/b(t))/b^2(t)$ has 
a scale-invariant form and is also known as an example
that the explicit form of the counterdiabatic term is
available~\cite{Jarzynski13, delCampo13}.

\section{Dynamical invariant and counterdiabatic driving}
\label{sec:app2}

The dynamical invariant is introduced as an auxiliary object 
to treat the solution of the Schr\"odinger equation
by an eigenvalue problem.
Once if we can find a pair of operators $I(t)$ and $H(t)$
satisfying Eq.~(\ref{di}), we can use it to construct
a counterdiabatic driving.

The counterdiabatic driving is formulated by introducing
an additional counterdiabatic term $H_1(t)$ for a given original
Hamiltonian $H_0(t)$~\cite{DR03, DR05, Berry09, Nakahara22}.
The total Hamiltonian is given by 
\begin{align}
 {\cal H}(t)= H_0(t)+H_1(t).
\end{align}
When the original Hamiltonian is written by the spectral representation 
\begin{align}
 H_0(t)=\sum_n E_n(t)|n(t)\rangle\langle n(t)|,
\end{align}
the counterdiabatic term is written as 
\begin{align}
  H_1(t)=i\hbar\sum_{m,n (m\ne n)}|m(t)\rangle
  \langle m(t)|\dot{n}(t)
  \rangle\langle n(t)|.
\end{align}

As a special case,
when the eigenvalues $E_1(t), E_2(t), \dots$ are independent of $t$,
we can identify 
$H_0(t)$ as a dynamical invariant and 
$H_1(t)$ as the corresponding Hamiltonian:
\begin{align}
  & H_0(t)=\epsilon I(t), \\
  & H_1(t)=H(t).
\end{align}
Since the dimension of the dynamical invariant is arbitrary,
we introduce a constant $\epsilon$ such that the dimension
of $\epsilon I(t)$ coincides with the dimension of energy.
When the eigenvalues of $H_0(t)$ are time-dependent,
$H_0(t)$ and $H_1(t)$ satisfy the relation 
\begin{align}
  [H_0(t),i\hbar\partial_t H_0(t)-[H_1(t),H_0(t)]]=0. \label{h01}
\end{align}
Thus, the dynamical invariant is interpreted as the special case
where the second entry in the commutation relation vanishes.
Equation (\ref{h01}) is recognized as a method for obtaining
approximate counterdiabatic terms~\cite{SP17, HT21}.
For a given $H_0(t)$, we seek $H_1(t)$ such that the norm of
the left hand side takes a minimum value.

\section{Another views of dynamical invariant}
\label{sec:app3}

As we mentioned before,
the equation for the dynamical invariant (\ref{di}) takes a familiar form.
For example, the Liouville--von Neumann equation takes the same form
as Eq.~(\ref{di}).
Then, the density operator represents a dynamical invariant.
This property shows that the dynamical invariant is not
a special quantity but is ubiquitous in any quantum systems.

In this section, we present several problems that use
an equivalent object to the dynamical invariant.
We expect that those examples offer another views
on the method of shortcuts to adiabaticity.

\subsection{Lax pair}

A relation between quantum shortcuts to adiabaticity
and classical nonlinear integrable systems was pointed out
in Ref.~\cite{OT16}.
In classical nonlinear integrable systems,
we treat nontrivial nonlinear equations.
The integrability denotes that the system
has infinite number of conserved quantities.
There are highly sophisticated techniques on such systems.
The Lax formalism was used to describe the integrable systems
in a unified way~\cite{Lax}.

A set of two operators $(L(t),M(t))$ is called a Lax pair
when it satisfies
\begin{align}
 \frac{\partial L(t)}{\partial t}=[M(t),L(t)].
\end{align}
We see from the comparison to Eq.~(\ref{di}) 
that $L$ is equivalent to the dynamical invariant.
The existence of the Lax pair represents the integrability of
the corresponding classical nonlinear system.
We can define $\psi$ satisfying 
\begin{align}
  & L\psi=\lambda\psi, \\
  & \frac{\partial}{\partial t}\psi=M\psi.
\end{align}
In the classical nonlinear integrable systems,
$\psi$ is used to construct the solution of
the corresponding nonlinear equation by the inverse scattering method.
The existence of the dynamical invariant implies
infinite series of conserved quantities represented by
the time-independent eigenvalues of $L$, $\lambda$. 
Instead of giving general discussions, 
we here introduce several examples
that are relevant to the present problems.

The most familiar example is given by
the following form of the Lax pair: 
\begin{align}
  & L=-\frac{\partial^2}{\partial x^2}+u(x,t), \\
  & M=-4\frac{\partial^3}{\partial x^3}
  +3\frac{\partial}{\partial x}u(x,t)
  +3u(x,t)\frac{\partial}{\partial x}.
\end{align}
Here, $u(x,t)$ is real and satisfies
the Korteweg-de Vries (KdV) equation~\cite{KdV}
\begin{align}
  \frac{\partial u(x,t)}{\partial t}=
  6u(x,t)\frac{\partial u(x,t)}{\partial x}
  -\frac{\partial^3u(x,t)}{\partial x^3}.
\end{align}
This equation is known to have multi-soliton solutions.
The simplest solution is the single soliton  
\begin{align}
  u(x,t)=\frac{-2\kappa^2}{\cosh^2(\kappa x-4\kappa^3t)},
\end{align}
where $\kappa$ is a positive constant.
This form of the Lax pair is practically useful since
$L$ is interpreted as a one-dimensional Hamiltonian
with a moving soliton potential.
To prevent the nonadiabatic transitions, we need to introduce
the counterdiabatic term obtained from the form of $M$.
It involves a cubic term in momentum operator
and is difficult to implement.
However, we can discuss a deformation of the counterdiabatic term
to a simple implementable form~\cite{OT16}.

The example of the KdV system 
is not a mere example of quantum controls.
It is known in classical nonlinear integrable systems
that the hierarchical structure exists in the KdV systems.
We can find an infinite series of Lax pairs.
This means that we can find the counterdiabatic terms exhaustively
in the type of a Hamiltonian $H=p^2+u(x,t)$.

As a promising example for physical implementations,
we present nonlinear lattice systems described by
the Toda equations~\cite{Toda1, Toda2}, which
can be represented by many-spin systems.
For $N$-qubit systems, the Lax pair is written as 
\begin{align}
  & L=\frac{1}{2}\sum_{n=1}^N J_n(t)\left(
  \sigma_n^x\sigma_{n+1}^x+\sigma_n^y\sigma_{n+1}^y\right)
  +\frac{1}{2}\sum_{n=1}^N h_n(t)\sigma_n^z, \\
  & M=-\frac{i}{2}\sum_{n=1}^N J_n(t)\left(
  \sigma_n^x\sigma_{n+1}^y-\sigma_n^y\sigma_{n+1}^x\right).
\end{align}
$J_n(t)$ and $h_n(t)$ satisfy the Toda equations
\begin{align}
  & \frac{dJ_n(t)}{dt}=J_n(t)(h_{n+1}(t)-h_n(t)), \\
  & \frac{dh_n(t)}{dt}=2(J_n^2(t)-J_{n-1}^2(t)).
\end{align}
It is known that the Toda equations also have multi-soliton solutions.
The corresponding spin Hamiltonian for $L$
is an isotropic XY model (XX model) with
a magnetic field in $z$-direction.
By using a solitonic form of the coupling constant $J_n(t)$
and the magnetic field $h_n(t)$, we can discuss a spin transport
in a spin-chain system~\cite{OT16}.
It is also known that the present spin model can be mapped onto
a fermion model by using the Jordan--Wigner transformation.
The coupling $J_n$ denotes a hopping between adjacent sites in that case.

\subsection{Quantum brachistochrone equation}

Various optimal control of systems may be obtained in 
shortcuts to adiabaticity.
The word ``optimal'' is somewhat ambiguous and
its meaning strongly depends on the problems to be solved.
we can consider optimizations in many different ways.
The method of quantum brachistochrone
is one of the optimization methods and can be a prominent method 
by its generality~\cite{CHKO06}.
For a quantum trajectory,
we define an action to determine the optimal Hamiltonian and
the corresponding state by a variational principle.
The main part of the action is determined from the geometric structure
of quantum states.
It is well known that the overlap between quantum states is characterized
by the Fubini-Study metric.

To solve practical optimization problems,
we introduce constraints for the Hamiltonian to be obtained.
Then, the minimization condition of the action gives
a form in Eq.~(\ref{di}).
For example, when we represent the constraints as 
\begin{align}
  {\rm Tr}\,H(t)X_\mu = h_\mu(t),
\end{align}
by using basis operators $X_\mu$ with $\mu=1,2,\dots,M$,
the corresponding dynamical invariant takes a form 
\begin{align}
  I(t)=\sum_{\mu=1}^M\lambda_\mu(t)X_\mu.
\end{align}
Here, $\{\lambda_n(t)\}$ is obtained by solving
the quantum brachistochrone equation.
It is not surprising that
the system is characterized by a dynamical invariant.
The point here is that the form of the dynamical invariant is
determined from the constraints.

The use of the action integral allows us to study stabilities
of the counterdiabatic driving~\cite{Takahashi13}.
The optimized solution is obtained by using the variation of the action
up to the first order.
The stability can be studied by expanding
the action up to the second order.

\subsection{Flow equation}

As a final application of the dynamical invariant,
we point out that the flow equation takes the same form
as the equation for the dynamical invariant.
The flow equation is a method diagonalizing a matrix by
iterations~\cite{GW93, GW94, Wegner94, Kehrein}.
For a given Hermitian matrix $H$, we consider
a time evolution described by 
\begin{align}
 i\frac{\partial H(t)}{\partial t}=[\eta(t),H(t)], \label{flow}
\end{align}
with $H(0)=H$.
$\eta(t)$ represents a generator of the time evolution.
One of possible choices is given by~\cite{Wegner94}
\begin{align}
 \eta(t)=i[H_{\rm diag}(t),H(t)],
\end{align}
where $H_{\rm diag}(t)$ is obtained by setting off-diagonal components
of $H(t)$ to zero.
That is, we have for a given set of base kets 
$\{|n\rangle\}$
\begin{align}
 \langle m|H_{\rm diag}(t)|n\rangle=\delta_{m,n}\langle n|H(t)|n\rangle.
\end{align}
Then, we can show
\begin{align}
  \frac{\partial}{\partial t}\sum_{m,n (m\ne n)}|\langle m|H(t)|n\rangle|^2
  &=-2\sum_{m,n}(\epsilon_n(t)-\epsilon_m(t))^2|\langle m|H(t)|n\rangle|^2
  \le 0, \label{negative}
\end{align}
where $\epsilon_n(t)=\langle n|H(t)|n\rangle$.
This relation shows that the magnitude of
each off-diagonal component of $H(t)$
gradually decreases as a function of $t$. 
Then, we expect 
\begin{align}
 \lim_{t\to\infty} H(t)=\lim_{t\to\infty} H_{\rm diag}(t).
\end{align}
Since Eq.~(\ref{flow}) is equivalent to Eq.~(\ref{di}),
the eigenvalues of $H(t)$ are independent of $t$, 
which means that the diagonal components at $t\to\infty$
represent the eigenvalues of the original matrix $H$.

From the aspect of shortcuts to adiabaticity,
the generator $\eta(t)$ is interpreted as a counterdiabatic term
for $H(t)$.
We specify the form of the counterdiabatic term
instead of specifying the time dependence of the matrix $H(t)$.
Then, the resulting dynamics is interpreted as
a diagonalization process of the original matrix $H$.

\section{Summary}
\label{sec:summary}

We have presented a brief introduction to the dynamical invariant
formalism of shortcuts to adiabaticity.
After some of fundamental properties are summarized,
we discussed the method of inverse engineering
together with several simple examples.
We also discussed the relation to the counterdiabatic driving
and several different aspects of the dynamical invariant.

The most important property of the dynamical invariant 
is that we can understand the dynamical system
in the same way as the static systems.
Once if we can find the dynamical invariant operator,
the problem is reduced to solving an eigenvalue equation.
For our purposes of quantum control, it is not necessary to
solve the eigenvalue problem.
Due to many possible choices of the coefficients of the dynamical invariant, 
the resulting protocol is not unique and
is obtained in an adapted manner.

Although experimental implementations
can be broadly covered by the examples of
a two-level system and a harmonic oscillator,
it is an interesting challenging problem
to find a dynamical invariant for other systems.
We expect that we can find a nontrivial use of
the dynamical invariant
by combining ideas from different fields
utilizing a similar quantity to the dynamical invariant.

\enlargethispage{20pt}





\funding{The author was supported by 
JSPS KAKENHI Grants No. JP20K03781 and No. JP20H01827. }

\ack{The author is grateful to Gonzalo Muga and Mikio Nakahara
for useful comments.}



\end{document}